# Deduction of the Bromilow's time-cost model from the fractal nature of activity networks

Alexei Vazquez*

Nodes & Links Ltd, Salisbury House, Station Road, Cambridge, England, CB1 2LA

*Email: alexei@nodeslinks.com

**Abstract**

In 1969 Bromilow observed that the time $T$ to execute a construction project follows a power law scaling with the project cost $C$, $T \sim C^B$ [Bromilow 1969]. While the Bromilow's time-cost model has been extensively tested using data for different countries and project types, there is no theoretical explanation for the algebraic scaling. Here I mathematically deduce the Bromilow's time-cost model from the fractal nature of activity networks. The Bromislow's exponent is $B=1-\alpha$, where $1-\alpha$ is the scaling exponent between the number of activities in the critical path $L$ and the number of activities $N$, $L \sim N^{1-\alpha}$ with $0 \leq \alpha < 1$ [Vazquez et al 2023]. I provide empirical data showing that projects with low serial/parallel (SP)% have lower $B$ values than those with higher SP%. I conclude that the Bromilow's time-cost model is a law of activity networks, the Bromilow's exponent is a network property and forecasting project duration from cost should be limited to projects with high SP%.

**1- Assumptions**

Let us consider a project with cost $C$, duration $T$, number of activities $N$ and critical path size of $L$ activities. The project cost is deduced from the sum of the cost of individual activities and, by the central limit theorem, it is approximated by

(1) $\quad C \approx c\,N$,

where $c$ is the average activity cost.

The project duration is deduced from the sum of critical path activities durations and, by the central limit theorem, it is approximated by

(2) $\quad T \approx t_c\,L$,

where $t_c$ is the average duration of critical path activities.

In [Vazquez et al 2023] it was demonstrated that there is a power law scaling between the critical path size and the number of activities

(3) $\quad L \approx A\,N^{1-\alpha}$,

where $A$ is a constant factor and $0 \leq \alpha < 1$ is an exponent that depends on the level of parallelism of the project activity network. For projects close to a linear chain of activities $\alpha \approx 0$ and $L \sim N$ as expected. As projects get parallelized $\alpha$ increases approaching the upper bound of $\alpha = 1$ for projects with almost all activities executed in parallel.

**2- Key result**

From (1) – (3) it follows that

(4) $\quad T \approx K\,C^B$,

with the Bromilow's exponent given by

(5) $\quad B = 1-\alpha$,

and the constant factor

(6) $\quad K = A\,t_c\,/\,c^B$.

*2.1- Implications*

1. Since $0 \leq \alpha < 1$ then $0 < B \leq 1$.
2. There is no unique value of *B* for all projects.
3. *B* is closer to 1 for projects with low parallelism, with few activities outside the critical path.
4. *B* is closer to 0 for projects with high parallelism, with several sub-critical paths.

**3- Empirical support**

*3.1- Data selection*

I have analyzed projects from the DSLIB database maintained by the Operations Research and Scheduling Research group at Ghent University https://www.projectmanagement.ugent.be/, downloaded on 2024-08-25. The project cards contain the Sector, reported Budget at Completion €, Planned duration Days and the Serial/Parallel (SP) %. The SP% is defined as

(7)  SP% = 100% (*L*-1) / (*N*-1) .

Vanhoucke *et al* 2008]. A total of 39 Construction Sector projects with no missing data and durations larger than 50 days were selected.

*3.2- Bromilow's exponent single project estimate*

According to equation (3), $\alpha$ should be estimated from the scaling between the critical path size and the number of activities. This can be done for simulated activity networks [Vazquez *et al* 2023], but it is not possible for single real projects. Yet, we can obtain a single-project estimate of $\alpha$ solving equation (3) for $\alpha$ and taking the limit of large critical path size ln*L*>>ln*A*, resulting in

(8)  $\alpha^* \approx \ln L / \ln N$ .

In turn, the critical path size can be calculated using equation (7), $L$ = (SP%/100%) (*N*-1) + 1. That allow us to obtain single-project estimates of $\alpha$. Bear in mind the resulting values are less precise for networks with a small critical path size.

The Bromilow's exponent *B* should be estimated from the plot of duration vs cost data. Yet, we can obtain a single-project estimate using $\alpha^*$ and the key result in equation (5), resulting in

(9)  $B^* = 1 - \alpha^*$.

The figure below shows that $\alpha^*$ decreases with increasing the SP% (Fig. 1A), while $B^*$ increases reaching almost 1 with increasing the SP% (Fig. 1B).

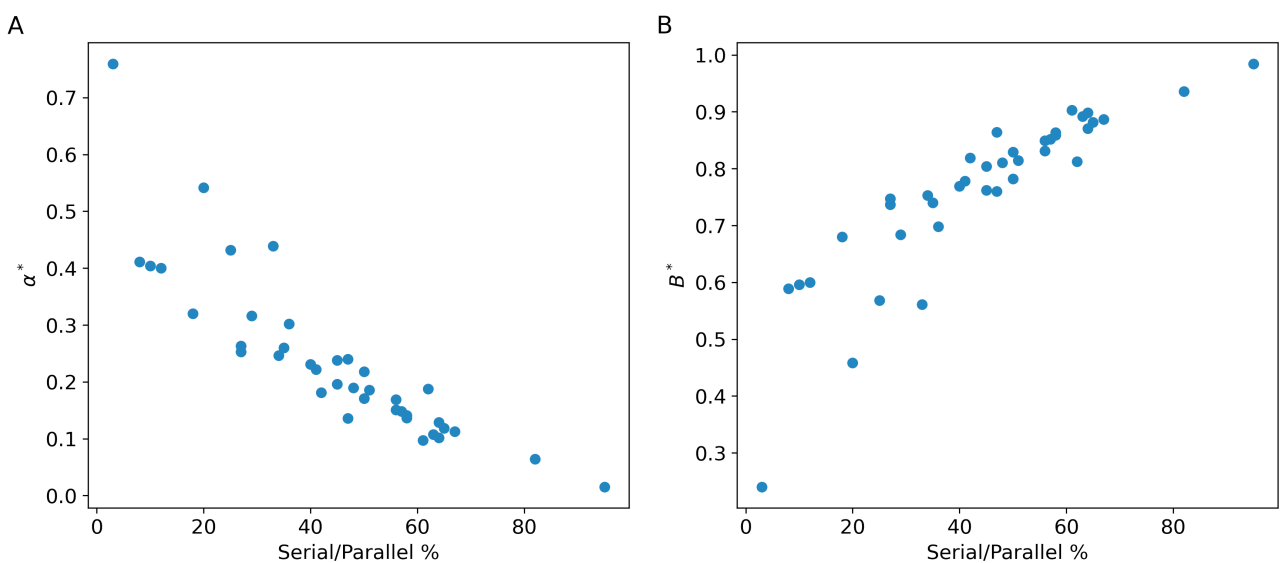

**Figure 1.** Bromilow's exponent estimation from single project data.

### 3.3- Bromilow's exponent multi project estimate

To provide further evidence that the Bromilow's exponent depends on the level of parallelism, I have divided the dataset into two quantiles with low and high SP%. Note that we're lumping together projects with different $B^*$, but that is the best we can do given the available data. Then I obtained an independent estimate of the Bromilow's exponent from the slope of a linear regression of log(Planned Duration Days) vs log(Budget at Completion €).

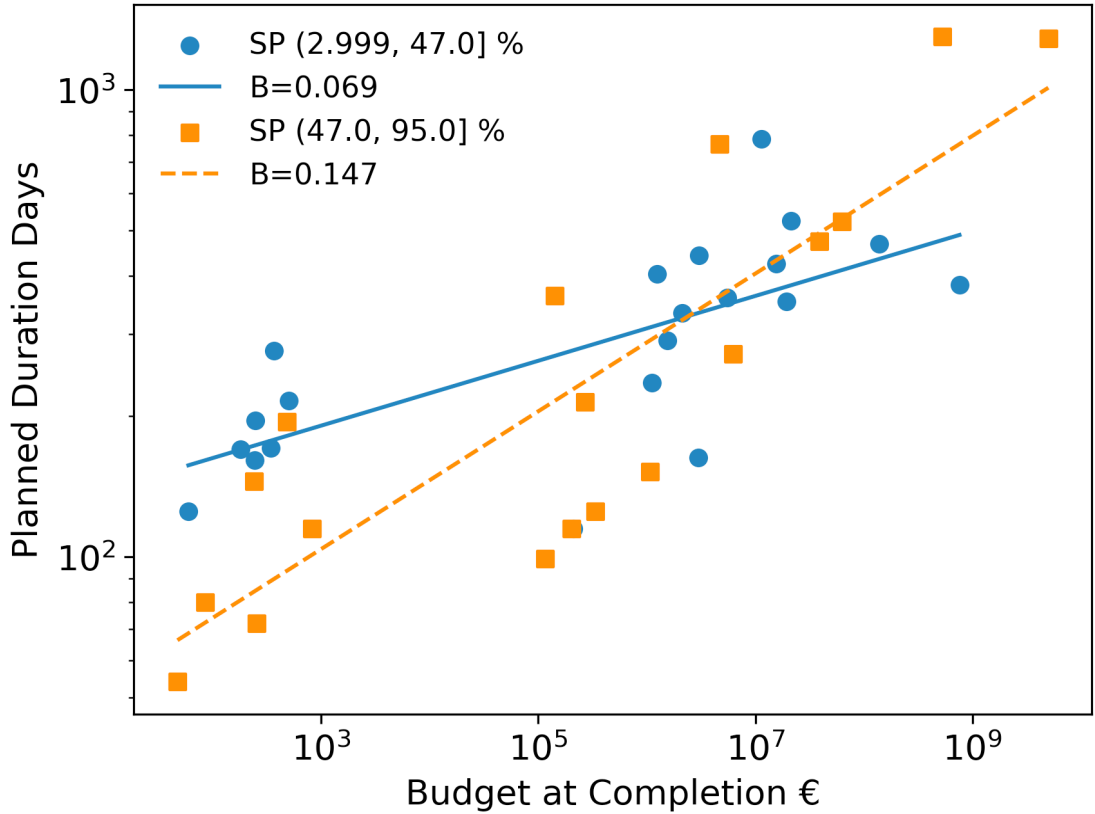

**Figure 2.** Bromilow's exponent estimation from duration vs budget data.

Despite similar budget ranges, projects with less parallelism (SP% (47.0, 95.0], Fig. 2 squares) have a wider range of durations than those with high parallelism (SP% (2.999, 47.0], Fig. 2 circles). In agreement with this observation, the Bromilow's exponent of the high SP% group is two times larger $B=0.147$, compared to $B=0.069$ for projects with low SP%. The chance to obtain a difference as large or larger is 0.0028 (100,000 permutations of the quantile labels) and therefore it is significant. This data supports the implications 1-4. The Bromilow's exponent is in the range $0<B\leq 1$, it is not unique for all projects and it is larger for projects with low parallelism (higher SP% quantile) than those with high parallelism (lower SP% quantile).

**Conclusions**

When deploying the Bromilow's model $T \approx K\,C^B$ to estimate project duration from cost, we should pay attention to the characteristics of the underlying activity network. The model parameters ($K$, $B$) should have been estimated using as input projects with similar level of parallelism to the target projects.

The prefactor $K$ is not an absolute constant (see equation (6)). The underlying assumption is that $C^B$ has larger variations across projects than $K$, and therefore the variations in $C^B$ determine the variations in project duration $T$. However, the Bromilow's exponent $B$ is small for projects with high level of parallelism (blue circles in Figure 2). In that context, the assumption that $C^B$ has larger variations across projects than $K$ does not hold true. Large variations in budget are not translated into high variations in project duration. I discourage the use of the Bromilow's model for projects with a Serial/Parallel below 50%.